\documentclass[onecolumn, journal, letterpaper,12pt]{IEEEtran}

\usepackage{graphicx, epstopdf}
\usepackage{amssymb, amsmath}
\usepackage{commath}
\usepackage{setspace}
\usepackage{acronym}
\usepackage{mathrsfs}
\usepackage{ifthen}
\usepackage{textcomp}
\usepackage{float}
\usepackage{subfigure}
\usepackage[table]{xcolor}
\usepackage{color,soul}
\usepackage{cite}
\usepackage{setspace}
\usepackage{mathtools}
\usepackage{bm}
\usepackage{url}
\usepackage{algorithm}
\usepackage{algpseudocode}
\usepackage[switch]{lineno}

\IEEEoverridecommandlockouts
\begin{document}

\title{Common Statistical Patterns in Urban Terrorism}

\doublespacing

\author{Weisi Guo\textsuperscript{1,2*}

\thanks{\textsuperscript{1}The Alan Turing Institute, United Kingdom. \textsuperscript{2}School of Engineering, University of Warwick, United Kingdom. }}

\maketitle

\begin{abstract}
The underlying reasons behind modern terrorism are seemingly complex and intangible. Despite diverse causal mechanisms, research has shown that there exists general statistical patterns at the global scale that can shed light on human confrontation behaviour. Whilst many policing and counter-terrorism operations are conducted at a city level, there has been a lack of research in building city-level resolution prediction engines based on statistical patterns.

For the first time, the paper shows that there exists general commonalities between global cities under terrorist attacks. By examining over 30,000 geo-tagged terrorism acts over 7000 cities worldwide from 2002 to today, the results shows the following. All cities experience attacks $A$ that are uncorrelated to the population and separated by a time interval $t$ that is negative exponentially distributed $\sim \exp(-A^{-1})$, with a death-toll per attack that follows a power law distribution. The prediction parameters yield a high confidence of explaining up to 87\% of the variations in frequency and 89\% in the death-toll data. These findings show that the aggregate statistical behaviour of terror attacks are seemingly random and memoryless for all global cities. 

The enabled the author to develop a data-driven city-specific prediction system and we quantify its information theoretic uncertainty and information loss. Further analysis show that there appears to be an increase in the uncertainty over the predictability of attacks, challenging our ability to develop effective counter-terrorism capabilities.
\end{abstract}

\section{Introduction}

Understanding complex human interactions is vital for solving some of humanity's most pressing social challenges \cite{Guo18}. One of these challenges is the protracted political violence that plague many urban regions in the world. Whilst creating data-driven regression models can yield insights into ongoing violence\cite{Hegre17,Hegre19}, statistical patterns can also yield insight into common trends \cite{Bohorquez09}. 

\subsection{Review of Statistical Analysis}
The science of finding patterns in war stretch back to the 1940s, when Richardson showed that the intensity of major battles in the Victorian era fits power-law distributions \cite{Richardson41, Denton68}. This has been reinforced for conflict and terrorism data in the modern era \cite{Bohorquez09, Larralde15}. Tracking trends in both large-scale wars and political violence is important to quantifying the effectiveness of peacekeeping and peace negotiation efforts \cite{Gleditsch05, owidwarandpeace, PRIO18}.

In terms temporal analysis (frequency or time interval between attacks), the frequency of large terrorist attacks have also been shown to obey a power-law distribution \cite{Clauset10, Clauset13}. Small-scale local events have also been studied, i.e. improvised explosive device (IED) attacks have been shown to exhibit self-excitation behaviour modeled by a Hawkes process \cite{Fry16}. Long term trend analysis has been conducted recently \cite{Clauset18}, whereby it is argued that our current period of relative peace from major wars is statistically insignificant.

The majority of statistical studies are still focused on aggregate scale attacks across a large region or the world \cite{Richardson41, Denton68, Clauset10, Clauset13, Clauset18, Bohorquez09, Larralde15}, scenario specific local attacks \cite{Fry16}, or long-term historical trends \cite{Dewey71} that span several centuries which follow various climate patterns~\cite{Hsiang11, Hsiang2013, Bai2011}. It remains an open question whether each city experiences \textit{a common human ecological behaviour in the frequency and size of attacks.} If so, it would inform urban policing and counter-terrorism policies and lead to\textbf{ city-specific prediction engines}. 

\subsection{Contribution}
Recent attempts have examined general behaviour at national statistical levels \cite{Mendes13} and at across different confrontation genres \cite{Johnson13}. However, detailed geographic analysis (city scale) across all genres and geographies is lacking. Indeed, city scale modeling is important as counter-terrorism policies are often adopted at the city scale (i.e., London and New York suffer disproportionately more threats and attacks than other cities) \cite{Graham_04}. Understanding a common ecological behaviour at detailed city resolution can help stakeholders to create models and make forecasts.

This paper sets out to do this. In this paper, the author offers insight that inter-relates the intensity, frequency, and prediction uncertainty of terrorist attacks in different cities worldwide since 2001. The results across different \textbf{urban ecologies} show that the \textbf{intensity} (death-toll per attack) data still obeys a power law \cite{Clauset10, Clauset13}, the \textbf{frequency} (interval between attacks) is exponentially distributed. This enables us to build a simple city-specific predictor based on past attack data, and we show both the information theoretic uncertainty and information loss in attempting to predict the underlying terrorism process. Finally, we use spectrogram analysis to further show that there is a growing uncertainty hidden in the complex process.

\section{Results}
\label{sec:results}

By analyzing geo-tagged terrorism and unconventional conflict data from the Global Terrorism Database (GTD) \cite{GTD}, the results show that the vast majority of conflict incidents occur in close proximity to an urban area with a mean distance of 27km. This highlights the importance of focusing on city-/town-scale resolution analysis. Fig.~\ref{fig:01}a shows a map of terrorist incidents (2002-14), where the black stars indicate the top 150 conflict locations (clustered to nearest city) with highest aggregate death-tolls. Fig.~\ref{fig:01}b-c shows example of data in the highest attacked city (Baghdad) with variations in death-toll and frequency as a function of time (day count).

\subsection{Intensity: Power Law Distribution}

\begin{figure}[t]
    \centering
    \includegraphics[width=1\linewidth]{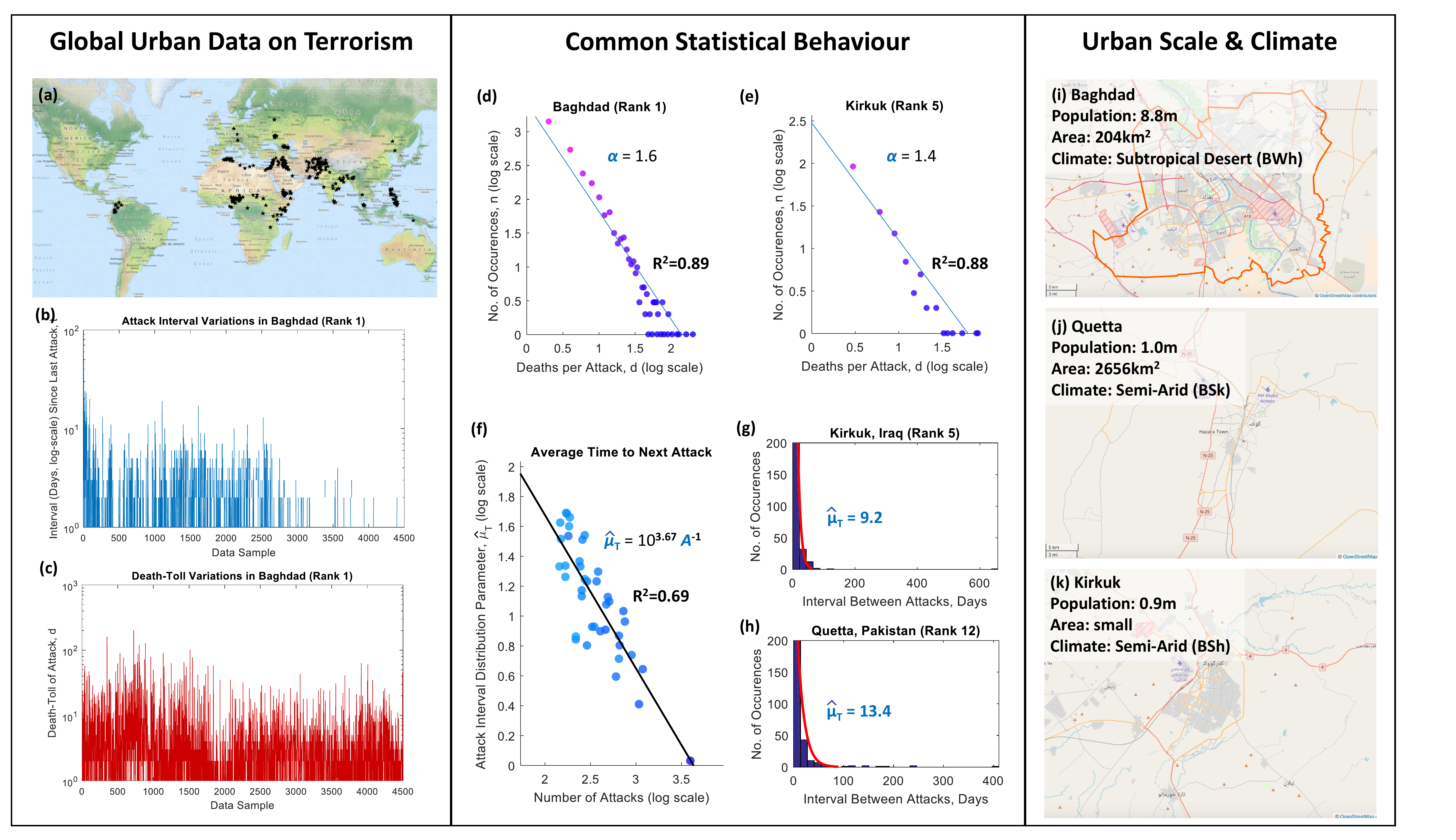}
    \caption{{\bf Terrorist Attack Intensity and Interval in Top 40 Conflict Cities with Diverse Urban Scales and Climates:} (\textbf{a}) Map of terrorist incidents (2002-14) - the black stars indicate the top 150 conflict locations (clustered to nearest city) with highest aggregate death-tolls. (\textbf{b-c}) Example of data in the highest attacked city (Baghdad) showing variations in death-toll and frequency as a function of time (day count). (\textbf{d-e}) the attack intensity follows power law distribution with exponent parameters $\alpha$ for two example cities. (\textbf{f}) the predicted average attack interval parameter $\hat{\mu}_T$ is linearly correlated with the average number of attacks $\hat{\mu}_T = 10^{3.67}/A$, where $10^{3.67}$ is the number of days in the recent 13 year interval. The actual number of attacks in each city explains for 69\% of the variation (adjusted $R^{2} = 0.69$) in the prediction parameter for all cities. (\textbf{g-h}) shows 2 example cities and the negative exponential distribution fit for interval (days) between attacks alongside the estimated parameter $\hat{\mu}_T$. (\textbf{i-k}) shows 3 example cities and their diverse population size (varies by 1 order of magnitude), diverse city area (varies by 1 order of magnitude) and different climates.}
    \label{fig:01}
\end{figure}

The results in Fig.~\ref{fig:01}d-e show that the terrorist intensity (death-toll per attack) is distributed in accordance to the established power law distribution \cite{Richardson41, Bohorquez09, Larralde15}. The exponent parameters $\alpha$ for two random cities are presented. The interesting observation is that most previous studies have considered low resolution conflicts (major wars) that span over 100 years, and it seems that the power law distribution remains valid even for high resolution terrorism and non-conventional conflict data in the modern era. What is less understood is how the time interval between attacks is distributed, and this is the focus of the paper.

\subsection{Interval: Negative Exponential Distribution}

The results show that the time interval between sequential attacks $t$ fit the negative exponential distribution. The pdf of a negative exponential distributed variable $t$ with support $[0,+\infty)$ is:
\begin{equation}
\label{exp}
\begin{split}
  f(t;\hat{\mu}_{T}) = \frac{1}{\hat{\mu}_{T}} \exp(-t/\hat{\mu}_{T}), 
\end{split}
\end{equation} where the parameter $\hat{\mu}_{T}$ is the distribution parameter. Note, the exponential decay rate is given by $1/\hat{\mu}_{T}$ and the variance is given by $\hat{\mu}_{T}^{2}$. Indeed, there are prior work to support this for civil war models that can be modelled using zero-inflated count models \cite{Bagozzi15}.

Fig.~\ref{fig:01}f-h shows the terrorist attack interval in the top 40 conflict cities. In general, all cities examined experience attacks that are separated by a time interval ($t$, days) that is negative exponentially distributed $\sim \exp(-\hat{\mu}_T)$. The results show 2 example cities and the negative exponential distribution fit for interval between attacks and the deaths per attack, alongside the estimated distribution parameters $\hat{\mu}_T$. Under the Maximum Likelihood (ML) estimator, the exponential distribution's parameter $\hat{\mu}_T$ is equal to the mean of the data $\mu$, i.e., $\hat{\mu}_T = \sum_i T_i / A$, where $T_i$ is the actual interval between any 2 attacks, and $A$ is the total number of attacks in the city over all time ($10^{3.67}$ days in 2002-14). Fig.~\ref{fig:01}f shows the predicted interval parameter $\hat{\mu}_T$ is linearly correlated with the average number of attacks $\hat{\mu}_T = 10^{3.67}/A$. The actual number of attacks in each city explains for 69\% of the variation (adjusted $R^{2} = 0.69$) in the distribution parameter $\hat{\mu}_T$. 

\begin{figure}[t]
    \centering
    \includegraphics[width=1\linewidth]{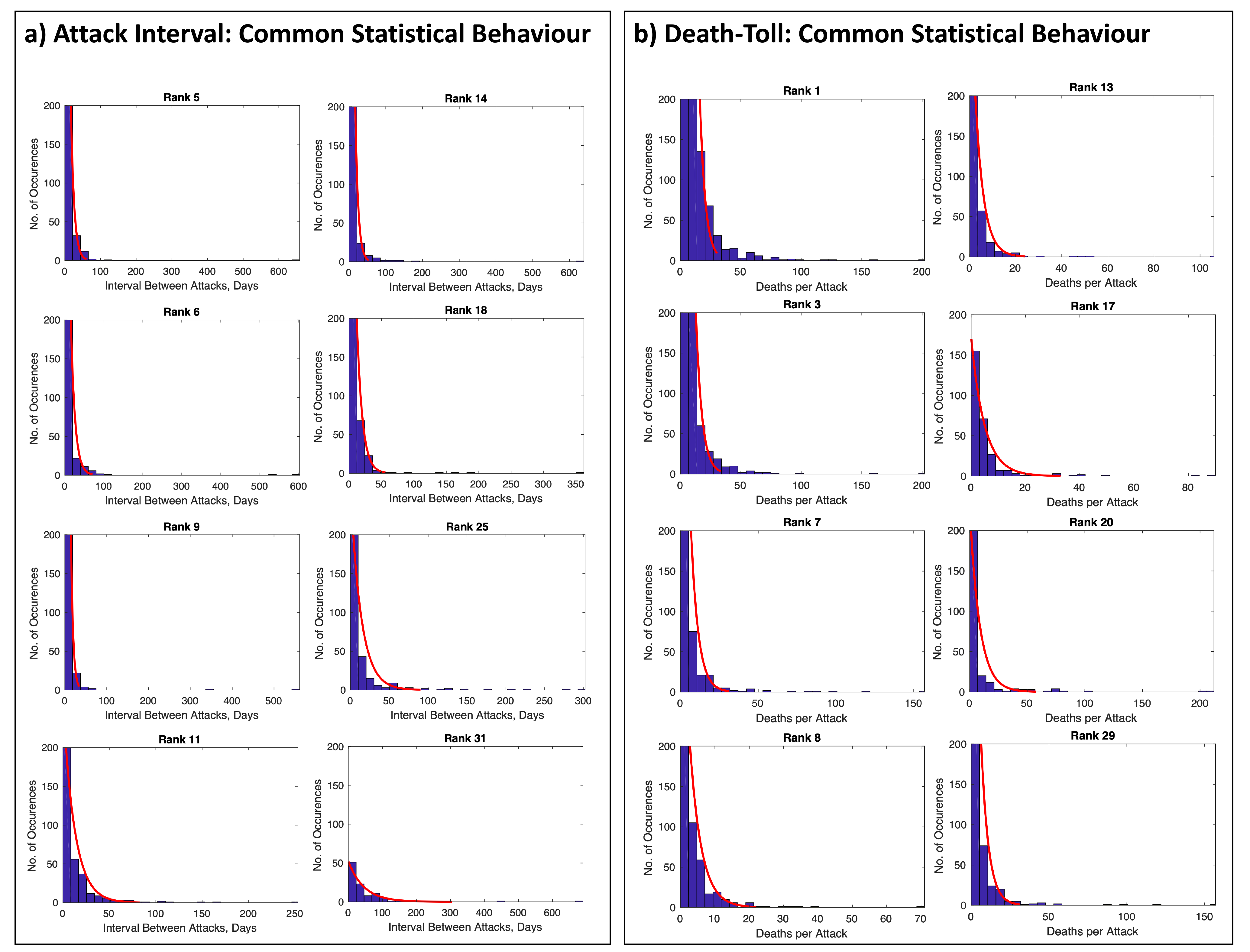}
    \caption{{\bf Terrorist Attack Interval and Intensity in Random Sample of Top 40 Conflict Cities with Diverse Urban Scales and Climates} (\textbf{a}) Common exponential distribution for attack intervals across random cities. (\textbf{b}) Common distribution for death-toll per attack across random cities.}
    \label{fig:A1}
\end{figure}

For a historical data sample of terrorist attacks that is $n$ in size, the lower- and upper-bound of the exponential distribution parameter is given as:
\begin{equation}
\label{exp_bound}
\begin{split}
  \hat{\mu}_{T,\text{upper}} &= \hat{\mu}_T \bigg( 1 - \frac{1.96}{\sqrt{n}}\bigg)^{-1} \\
  \hat{\mu}_{T,\text{lower}} &= \hat{\mu}_T \bigg( 1 + \frac{1.96}{\sqrt{n}}\bigg)^{-1}. 
\end{split}
\end{equation} For the top 40 cities considered in the analysis, the number of attacks between 2002-2014 is between 3983 (rank 1 conflict city) to 141 (rank 40), which yields percentage changes of 3\% and 14-19\% to the distribution parameter. This shows that the distribution given for the attack intensity and frequency is robust across different urban scales and climates. Fig.~\ref{fig:01}i-k shows 3 example cities and their diverse population size (varies by 1 order of magnitude), diverse city area (varies by 1 order of magnitude) and different climates. A large comparison set of random cities in the top 40 conflict cities is given in Fig.~\ref{fig:A1} which shows a common statistical distribution across all of them.
\begin{figure*}[t]
    \centering
    \includegraphics[width=1\linewidth]{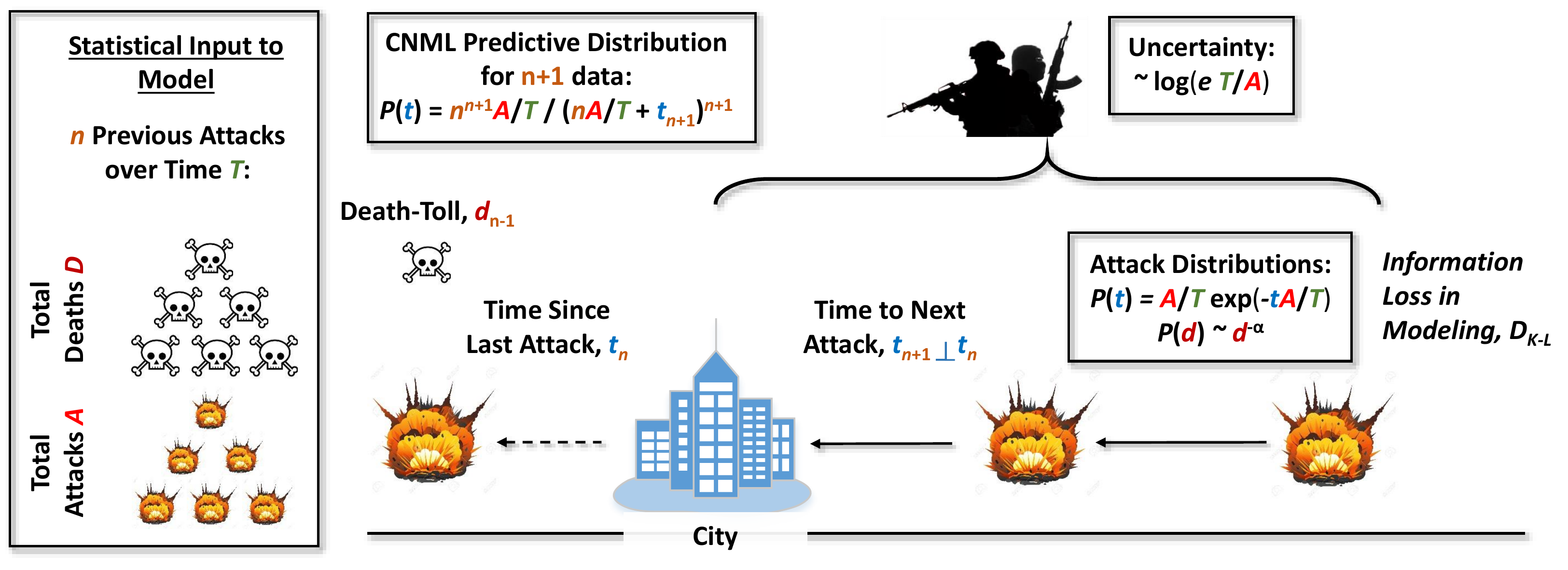}
    \caption{{\bf Prediction of Next Terrorist Attack in City:} the time to next attack $t_i$ is independent of the time since previous attack $t_{i-1}$, and in general the time between attacks is negative exponential distributed with dependency on the mean number of attacks $\sim \exp(-A/T)$. The death-toll per attack (intensity) is power law distributed. The uncertainty of the underlying terrorism process is logarithmically proportional to the mean time between attacks $\sim \log(e T/A)$. The only input parameters into the model are the number of attacks $A$ and deaths $D$ aggregated over a period of $T$.}
    \label{fig:5}
\end{figure*}

\subsection{Prediction Accuracy and Information Loss}

Exponential distributions are commonly associated with waiting time between random and memoryless events (i.e., Poisson point processes). Therefore, the fitted negative exponential distributions in Fig.~\ref{fig:01} indicate that sequential attacks in each city are unrelated. A possible reason is that each terrorist attack depends on a large number of variables (i.e., organization, logistics, finance, personnel, evading detection, and opportunity), which suppresses any dependency between attacks. One other interesting property means that the probability of the waiting time for the next attack is constant, irrespective of how much time has surpassed. That is to say, one only needs to understand one single parameter $\hat{\mu}_T$ in order to predict the time for the next attack, irrespective of when the last attack was. This is similarly true for the death-toll per attack: the number of deaths in the next attack is independent of the previous attack event's death-toll. 
\begin{figure}[t]
    \centering
    \includegraphics[width=0.75\linewidth]{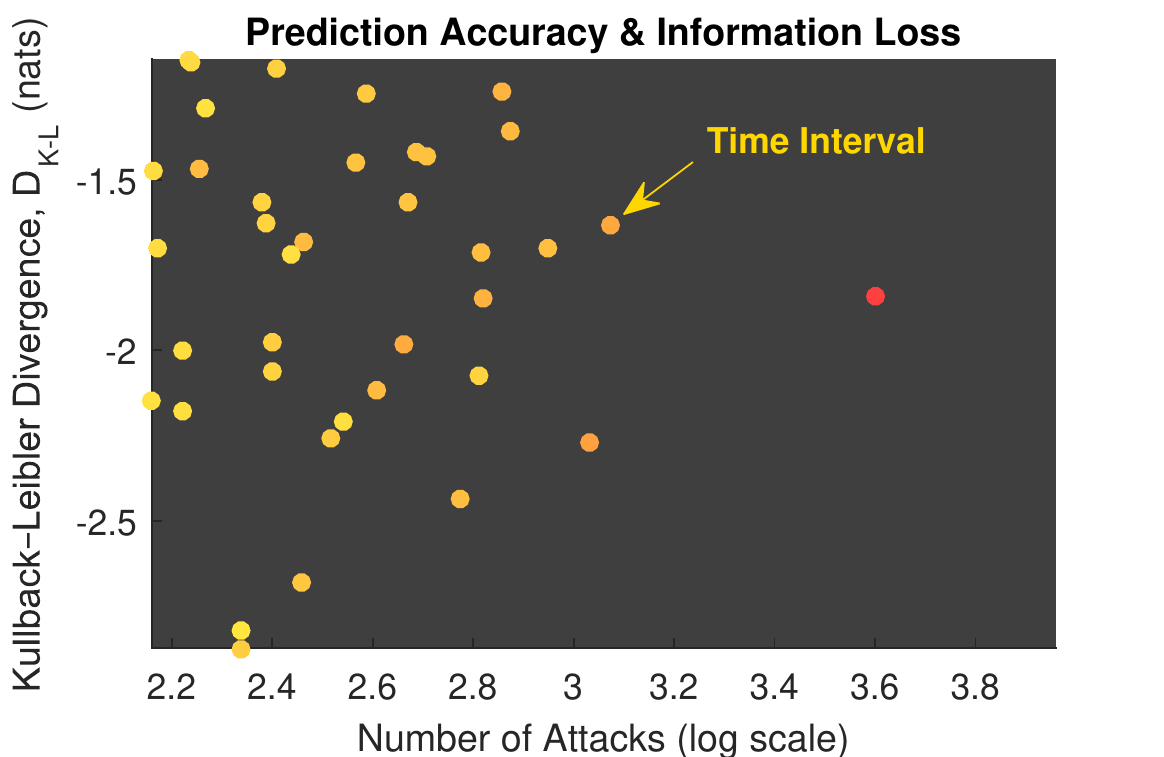}
    \caption{{\bf Accuracy and Information Loss in Prediction:} directed K-L divergence (information loss) for predicting the time interval between attacks.}
    \label{fig:2}
\end{figure}

As an example, Fig.~\ref{fig:5} illustrates a system for prediction of the next terrorist attack and the uncertainty of the terrorism process. The time to next attack $t_i$ is independent of the time since previous attack $t_{i-1}$, and in general the time $t$ between attacks is negative exponential distributed with dependency on the mean number of attacks $t \sim \exp(-A/T)$ and the death $d$ is power law with dependency on the exponent $\alpha$. The uncertainty of the underlying terrorism process is logarithmically proportional to the mean time between attacks $\log(e T/A)$. The only input parameters into the model are the number of attacks $A$ and deaths $D$ aggregated over a period of $T$. Among all continuous probability distributions with support $[0, +\infty)$ and mean $\hat{\mu}$, the exponential distribution has the largest entropy of $\log(e\hat{\mu})$: 
\begin{equation}
\label{Entropy_exp}
\begin{split}
  h(X) =& -\int_{0}^{+\infty} \frac{1}{\hat{\mu}}\exp(-x/\hat{\mu}) \log\bigg[\frac{1}{\hat{\mu}}\exp(-x/\hat{\mu})\bigg] \dif x \\
  =& 1 - \log(1/\hat{\mu}) = \log(e\hat{\mu}).
\end{split}
\end{equation}
This indicates a logarithmic higher information content (uncertainty) in the underlying terrorism processes (i.e., the terrorist organizations) that plan attacks with high waiting duration $\hat{\mu}_T$. 
\begin{figure*}[t]
    \centering
    \includegraphics[width=1.00\linewidth]{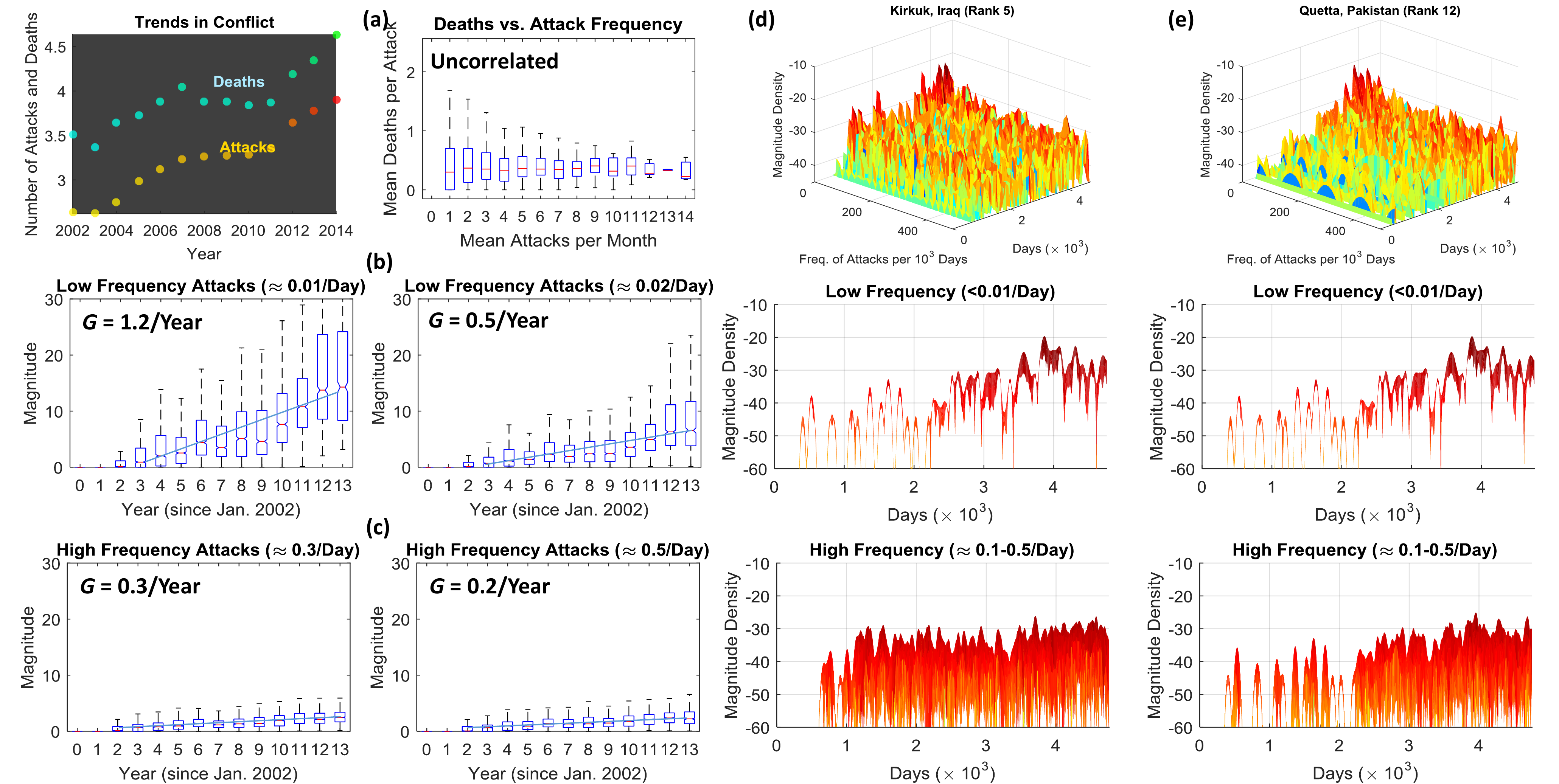}
    \caption{{\bf Temporal Variations in the Frequency of Attacks:} (\textbf{a}) the growing number of attacks and deaths and the death per attack is uncorrelated with the attack frequency. (\textbf{b}) the rapid growth ($G = 1.2-0.5$) in the magnitude of low frequency attacks. (\textbf{c}) the slow growth ($G = 0.3-0.2$) in the magnitude of high frequency attacks. (\textbf{d-e}) the spectrogram of the attacks between 2012-2014 for 2 example cities, with slices of low and high frequency magnitude variations as a function of time (days).}
    \label{fig:3}
\end{figure*}

In prediction, we assume that for an observed city that has suffered $n$ previous attacks and these attacks have a frequency and intensity that are both exponentially distributed with unknown parameters. A common predictor for random and negative exponentially distributed data $x$ is to use is the Maximum-Likelihood (ML) predictor, which yields the following predictive density: $p_{\text{ML}}(x_{n+1}|x_1,...,x_n) = \frac{1}{\mu}\exp(-\frac{x_{n+1}}{\mu})$, where $x_{n+1}$ is the future data value. An improved predictive distribution free of the issues of choosing priors is the Conditional Normalized Maximum Likelihood (CNML) estimator yields the following predictive density of the future data \cite{Schmidt09}:
\begin{equation}
\label{CNML}
\begin{split}
  p_{\text{CNML}}(x_{n+1}|x_1,...,x_n) = \frac{n^{n+1}\mu^{n}}{(n\mu+x_{n+1})^{n+1}},
\end{split}
\end{equation} where $\mu$ is taken from the data ($\{x_1,...,x_n\}$). 

Given that both the time interval and the intensity (death-toll) fit negative exponential distributions, based on Eq.\eqref{KL}, the directed Kullback-Leibler (K-L) divergence (information loss) of adopting an exponential model instead of using the data is:
\begin{equation}
\label{KL_exp}
\begin{split}
  D_{\text{K-L}}(\mu\|\hat{\mu}) = \log(\mu^{-1}) - \log(\hat{\mu}^{-1}) + \frac{\mu}{\hat{\mu}} - 1, 
\end{split}
\end{equation} where $\mu$ is the mean of the data and $\hat{\mu}$ is the parameter of the regression or predictor. In general, as shown in Fig.~\ref{fig:2}, the negative exponential model causes a loss in information that varies between 1.2-2.8 nats (0.7-3.3 bits). This demonstrates that despite the seemingly accurate statistically characterization of conflict frequency, there is a non-negligible surprise element from an information theoretic sense.

\subsection{Spectrogram Analysis}

The frequency and intensity of violence has shifted over the past decade and this will affect the long term accuracy of the proposed prediction model. Spectral analysis has the potential to observe the different frequency components of attacks and how they shift with time (2002-2014). As a hypothesis, it can potentially distinguish low frequency (long time interval $T/A$) high casualty (death-toll $D/A$) attacks from high frequency low casualty attacks. Spectral analysis using Short-Time-Fourier-Transform (STFT) is used on each city to produce spectrogram plots. The parameters used are Hamming window of size 128 (days), with non-overlap size of 120, 128 Fast-Fourier-Transform sampling points to calculate the Discrete Fourier Transform. In general, the STFT of a discrete sequence $x[n]$ is defined as: $X(m,\omega) = \sum_{-\infty}^{+\infty} x[n]w[n-m]\exp(-j\omega n)$, where $w[n]$ is the window function and $\omega$ is the continuous frequency, and the spectrogram is defined as $|X(m,\omega)|^{2}$. 

In Fig.~\ref{fig:3}, the results show the magnitude of the frequency of attacks as a function of time. Fig.~\ref{fig:3}(\textbf{a}-left) shows the growing number of attacks and show in (\textbf{a}-right) that the deaths per attack is uncorrelated with the attack frequency. Digging deeper using spectrogram analysis, the author shows that there is a rapid growth ($G = 1.2-0.5$) in the magnitude of low frequency attacks (\textbf{b}), and a slow growth ($G = 0.3-0.2$) in the magnitude of high frequency attacks (\textbf{c}). The results indicate that the growth in the number of attacks and deaths is largely due to low frequency attacks, which is causing a disproportionate number of high deaths (see the high variance in (\textbf{a}-right)). Fig.~\ref{fig:3}(\textbf{d-e}) show the spectrogram of the attacks for 2 example cities, with slices of low and high frequency magnitude variations as a function of time (days). The growing number of attacks (and deaths) is not due to population increases, as shown by the lack of correlation in Fig.~\ref{fig:A2} in the \textbf{Methods}.

In summary, the spectrogram analysis reveals that the growth in death-tolls from terrorist attacks seems to be due to an increase in the number of slower (1 attack per 100 days) and bigger casualty attacks (over 10 deaths per attack). Referring back to the entropy of the terrorism process (i.e., entropy is $\log(e T/A)$), the growing number of low-frequency (large $T/A$) and high death-toll indicates that the underlying terrorism process and organization is increasing in uncertainty. It is unclear how to combine the entropy measures if the high death-toll attacks are dependent on the longer planning process, and this is the focus of future research.

\section{Discussions and Conclusion}

Despite the seemingly complex reasons that drive modern terrorism, conflict and violence, this paper has shown that all modern conflicts exhibit common frequency and intensity patterns that can be modeled accurately to give true predictive powers to smart city systems. By examining over 30,000 geo-tagged conflict data points over 13 recent years, the paper demonstrates the following. The number of attacks and the death-toll is uncorrelated to the population of the city. The attacks are separated by a time that is negative exponentially distributed $\sim \exp(-\hat{\mu}_T)$ and the number of deaths per attack follows power law distributions. The prediction parameters explains for 69-87\% of the variations in real data. Whilst the parameters of the distributions vary between cities and with time, these findings show that the frequency of terror attacks is random and memoryless. 

The distributions found in this paper can be used to predict the next attack and the Kullback-Leibler divergence is used to show that approximately 0.7-3.3 bits of information is lost through the predictions. As such, future work should focus on integrating generalized statistical models presented in this paper with microscopic excitation and mechanical models \cite{Fry16}. 

Using spectrogram analysis, it was uncovered that the growth in death-tolls is due to a growing number of slower but higher casualty attacks. Combining the spectrogram analysis with the entropy analysis, the combined results seem to indicate a logarithmically increasing uncertainty in the underlying terrorism random process. This uncertainty makes prediction and developing counter-terrorism strategies more challenging. \\

\section{Methodology}
\label{sec:methods}

\subsection{Data}
\begin{figure}[t]
    \centering
    \includegraphics[width=0.75\linewidth]{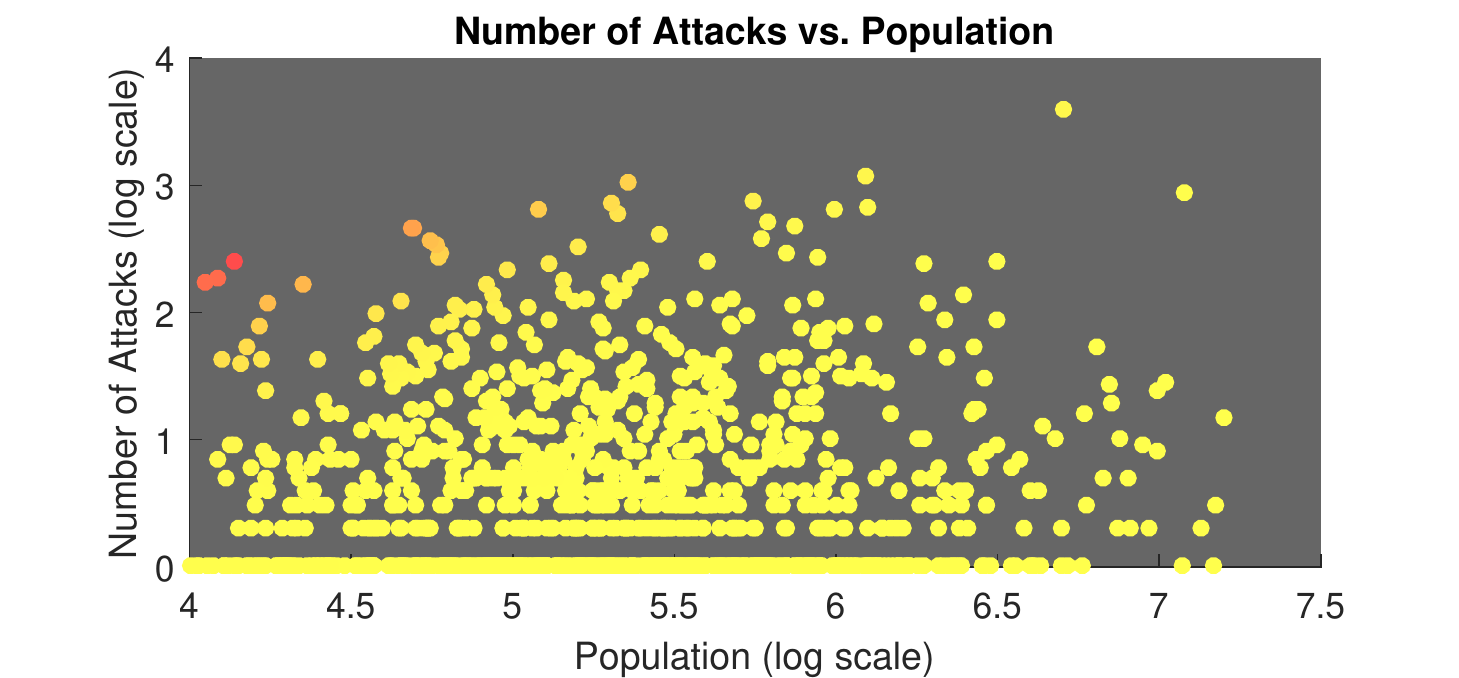}
    \caption{{\bf Attacks are Uncorrelated with the Population of the City.} }
    \label{fig:A2}
\end{figure}

The terrorism and conflict data used in this paper is sourced from 30,000+ attacks between 2002 to 2014 (13 years) from the Global Terrorism Database (GTD) \cite{GTD}. For each city and over time period $T$, the GTD contains the number of geo-tagged attacks $A$ and death-toll $D$ from incidents, which range from small-scale assassinations (1 death) to large-scale massacres (1000s dead). A plot of the major terrorist and conflict incidents is shown in Fig.~\ref{fig:01}a. The GTD data is then clustered to the nearest city\footnote{over 7000 cities and settlements were considered sourced from the National Geospatial Intelligence Agency \cite{Cities_Data}.}. As a result of clustering, the author shows that the vast majority of conflict incidents occur in close proximity to an urban area with a mean clustering distance of 27km. This means whilst most data points are in cities, some do occur in rural and suburban areas, which is still relevant from a policy perspective. It is worth noting that the number of attacks (and deaths) is not due to population increases, as shown by the lack of correlation in Fig.~\ref{fig:A2}. Therefore, models with predictive power are needed to understand the frequency and intensity of attacks for each city. 

Using the data, 2 variables are extracted: (1) the time interval $t$ between each consecutive attack (frequency), and (2) the death-toll per attack $d$ (intensity). Of the data obtained between 2002 and 2014, only 40 cities in the world have sufficient conflict data to obtain distributions from which the error in Maximum Likelihood (ML) parameter estimation is less than 20\% (see \textbf{Results} section). These cities range from the Middle East, West Africa, South Asia, to the Far East. A list of the cities can be found in Table~\ref{Top40}.  

\subsection{Metrics}
In order to compare between data sets, the coefficient of determination $R^{2}$ is used. It is a number that indicates how well the statistical regression model fits the data. In other words, the percentage of variance in the data that can be explained by the proposed model. For a data vector $y = [y_1, y_2, ... y_K]$ (with mean $\overline{y}$) and a predicted data vector using the regression model $\hat{y}$, the residue vector is defined as $e = y - \hat{y}$. The coefficient of determination $R^{2}$ is defined as $R^{2} \equiv 1 - \frac{\sum_{k}e_{k}^{2}} {\sum_{k}(y_{k}-\overline{y})^{2}}$, where the numerator is known as the residual sum of squares and the denominator is known as the total sum of squares. In this paper, the analysis employs the adjusted $R^{2}$ to take discount against extra variables in the model $\text{adjusted } R^{2} = 1 - (1-R^{2})\frac{K-1}{K-V-1}$, where $V$ is the number of variables in the regression model.

In order to compare between different probability distributions $P$ (true data) and $Q$ (regression model), the directed information gain/loss metric is used. The Kullback-Leibler (K-L) divergence of $Q$ from $P$ is $D_{\text{K-L}}(P\|Q)$, and it is defined as \cite{Cover_91}:
\begin{equation}
\label{KL}
\begin{split}
  D_{\text{K-L}}(P\|Q) = \int_{-\infty}^{+\infty} p(x) \log \frac{p(x)}{q(x)} \dif x, 
\end{split}
\end{equation} where $p(x)$ and $q(x)$ denote the densities of $P$ and $Q$. For exponential distributions, this is given as: $D_{\text{K-L}}(\mu\|\hat{\mu}) = \log(\mu^{-1}) - \log(\hat{\mu}^{-1}) + \frac{\mu}{\hat{\mu}} - 1$.  

\subsection{Supporting Analysis}

\begin{table}[ht!]
    \caption{Top 40 prominent conflict cities}
    \begin{center}
        \begin{tabular}{|l|l|l|l|}
          \hline
          \emph{City(1-20)} & \emph{Country}    &\emph{City(21-40)}& \emph{Country}   \\   
          \hline
          Baghdad           & Iraq              &Musayyib          & Iraq              \\
          Mosul             & Iraq              &Banghazi          & Libya             \\
          Baqubah           & Iraq              &Farah             & Afghanistan       \\
          Karachi           & Pakistan          &Zareh             & Afghanistan       \\
          Kirkuk            & Iraq              &Jalalabad         & Afghanistan       \\
          Lashkar           & Afghanistan       &Kabul             & Afghanistan       \\
          Peshawar          & Pakistan          &Ghazni            & Afghanistan       \\
          Mogadishu         & Somalia           &Saidu             & Pakistan          \\
          Yala              & Thailand          &Kohat             & Pakistan          \\
          Fallujah          & Iraq              &Pattani           & Thailand          \\
          Kandahar          & Afghanistan       &Qalat             & Afghanistan       \\
          Quetta            & Pakistan          &Cotabato          & Philippines       \\
          Asadabad          & Afghanistan       &Tall Afar         & Iraq              \\
          Tikrit            & Iraq              &Meymaneh          & Afghanistan       \\
          Ramadi            & Iraq              &Qasr Shirin       & Iran              \\
          Bannu             & Pakistan          &Gardiz            & Afghanistan       \\
          Parachinar        & Pakistan          &Baraki            & Afghanistan       \\
          Narathiwat        & Thailand          &Sukkur            & Pakistan          \\
          Samarra           & Iraq              &Groznyy           & Russia            \\
          Maiduguri         & Nigeria           &Tizi-Ouzou        & Algeria           \\
          \hline
        \end{tabular}
    \end{center}
    \label{Top40}
\end{table}

\pagebreak

\end{document}